\documentclass[a4paper]{jpconf}
\usepackage{graphicx}
\usepackage{wrapfig}
\begin{document}
\title{Recent results from the STAR spin program at RHIC}

\author{M. Sarsour for the STAR Collaboration}

\address{Cyclotron Institute, Texas A$\&$M University, College Station,
 Texas 77843, USA}

\ead{msar@tamu.edu}

\begin{abstract}
  The STAR experiment uses polarized p+p collisions at RHIC to determine
 the contributions to the spin of the proton from gluon spin and from
 orbital angular momentum of the quarks and gluons. Selective STAR
 measurements of the longitudinal double spin asymmetry for inclusive jet
 and inclusive hadron production are presented here. In addition, we
 report measurements of the transverse spin asymmetry for di-jet
 production at mid-rapidity and the transverse single-spin asymmetry for
 forward $\pi^\circ$ production.
\end{abstract}

\medskip

  The quark and anti-quark contributions to the spin of the proton
 measured with polarized deep-inelastic scattering (DIS) fixed target
 experiments amount to only $\le30\%$ \cite{ANP26-1}. Consequently, the
 rest of the spin of the proton must come from the
 gluons and the orbital angular momentum of both quarks and gluons.
 However, the gluon polarization, $\Delta g$, is poorly constrained from
 scaling violations in the DIS data \cite{PRD53-6100}, and more precise
 and direct measurements are needed. Polarized p+p collisions at the
 Relativistic Heavy Ion Collider (RHIC) provide a very suitable
 environment, rich with strongly interacting probes, to measure $\Delta g$
 directly and precisely. There are many processes where the gluon
 participates directly. In addition, the high c.m. energy, $\sqrt{s}$ =
 200 GeV, and high $p_T$ make next-to-leading order (NLO) perturbative
 Quantum Chromodynamics (pQCD) analysis more reliable. The spin program of
 the Solenoid Tracker at RHIC (STAR) experiment \cite{NIMA499-624} aims,
 in the short term, to utilize these advantages to measure $\Delta g$
 using inclusive jet and inclusive hadron production measurements in
 longitudinally polarized p+p collisions. In addition, the STAR experiment
 uses transversely polarized p+p collisions to gain insights into the
 orbital angular momentum of the partons.

\medskip
\medskip

\begin{figure}[h]
 \begin{minipage}{10pc}
 \includegraphics[scale=0.29]{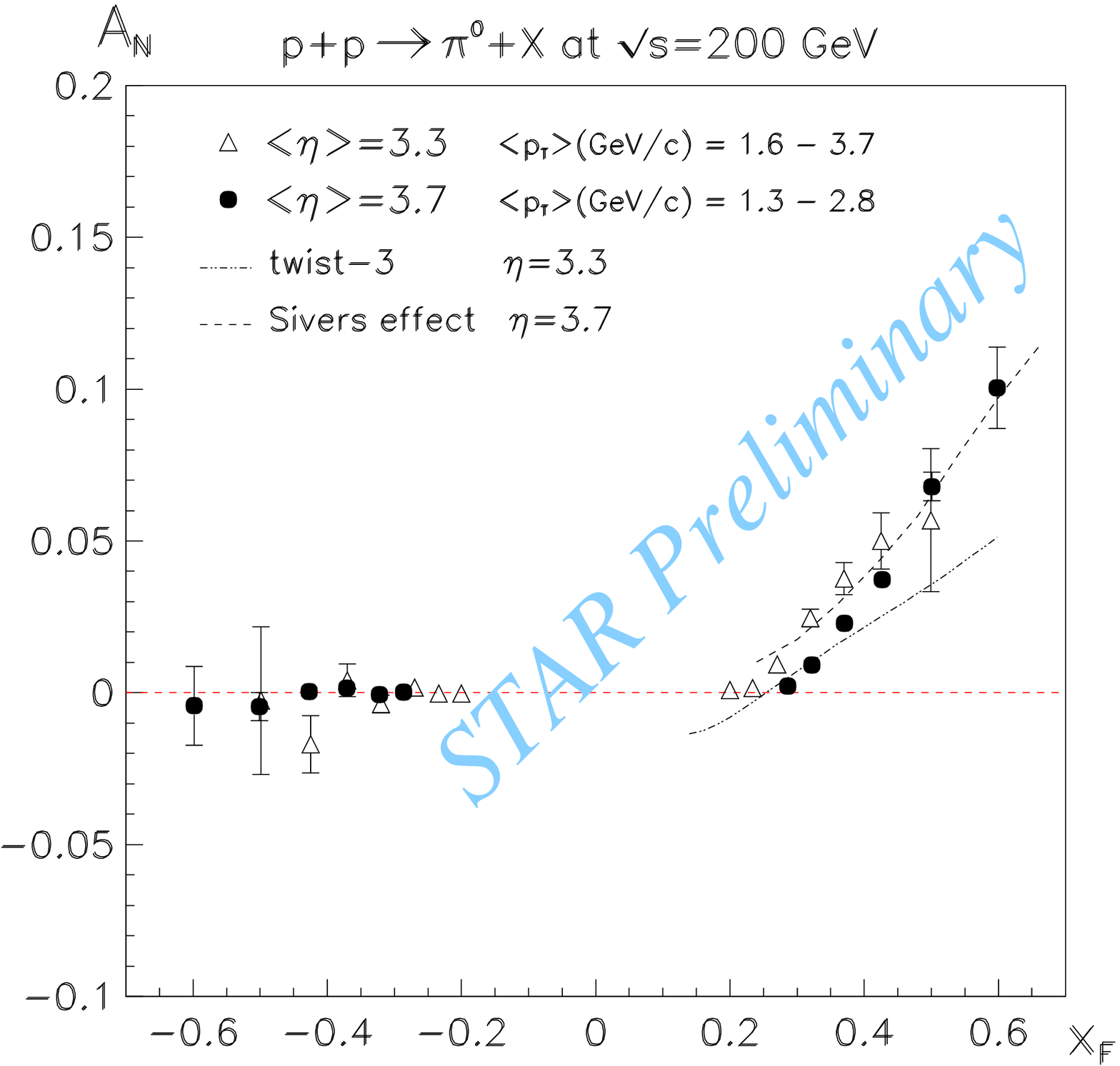}
 \end{minipage}\hspace{3pc} 
 \begin{minipage}{10pc}
 \includegraphics[scale=0.47]{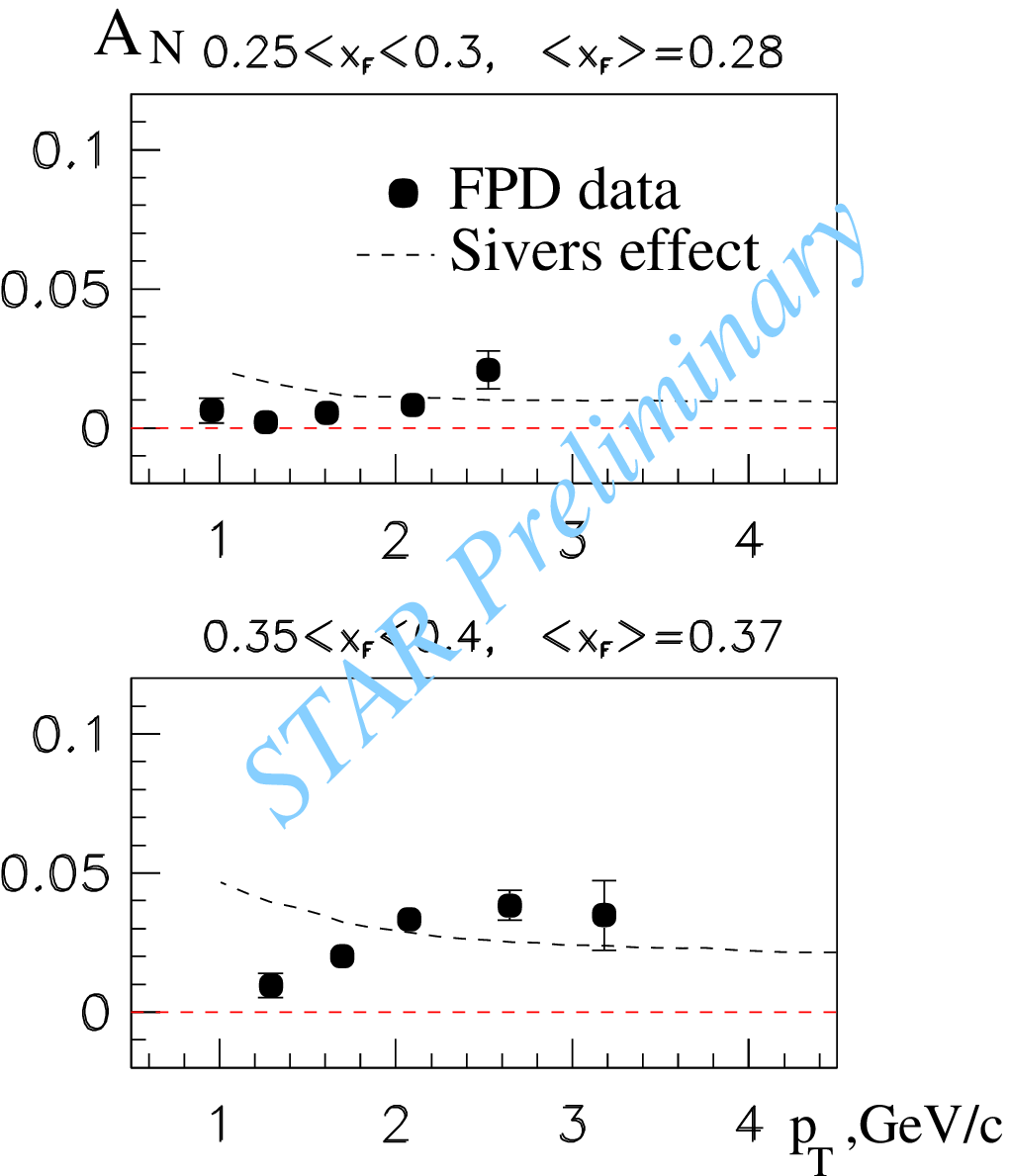}
 \end{minipage}\hspace{4pc}
 \begin{minipage}{14pc}
 \parbox[h]{4.4cm}{\caption{\label{fig:anvxfnpt}\small Left panel: $A_N$
 as a function of $x_F$ at $\langle\eta\rangle = 3.3$ (triangles) and
 $\langle\eta\rangle = 3.7$ (circles) with statistical uncertainties. The
 lines are Twist-3 \cite{hepph-0609238} and Sivers effect
 \cite{PRD74-094011} based calculations. Right panel: $A_N$ as a function
 of $p_T$ for different $x_F$ bins along with Sivers effect based
 calculations .}}
 \end{minipage}
 \end{figure}

  The STAR collaboration previously observed a sizable analyzing power,
 $A_N$, for forward $\pi^\circ$ production at large $x_F$
 \cite{PRL94-171801}, similar to previous observations at lower energies
 \cite{PLB264-462}. The previous STAR results were in qualitative
 agreement with several different model predictions \cite{PRL94-171801}
 and could not differentiate between them, so higher precision
 measurements of $A_N$ as a function of both $x_F$ and $p_T$ were needed.
 New $A_N$ measurements with higher luminosity and polarization were
 performed in 2006 \cite{hepex-0612030}. Figure~\ref{fig:anvxfnpt} shows
 preliminary results for $A_N$ as a function of $x_F$ (left panel) and as
 a function of $p_T$ for different $x_F$ bins (right panel), along with
 model predictions.
 The results in the left panel are consistent with the previous results
 \cite{PRL94-171801} in having large $A_N$ for high $x_F$, and provide
 sufficient accuracy to discriminate between different dynamics. The right
 panel shows that the $p_T$ dependence of $A_N$ is not well described
 by Sivers based calculations \cite{PRD74-094011,hepph-0609238}, which
 predict that $A_N$ should decrease as you go higher in $p_T$ for all
 $x_F$ bins.

\medskip

 The STAR experiment also probed the Sivers effect by measuring the di-jet
 opening angle in transversely polarized p+p collisions \cite{janProc}
 since Boer and
 Vogelsang \cite{PRD69-094025} suggested that $A_N$ for the distribution
 in relative azimuthal angle, $\zeta$, between the di-jet pairs could
 provide
 \begin{wrapfigure}{r}{95mm}
 \includegraphics[scale=0.5]{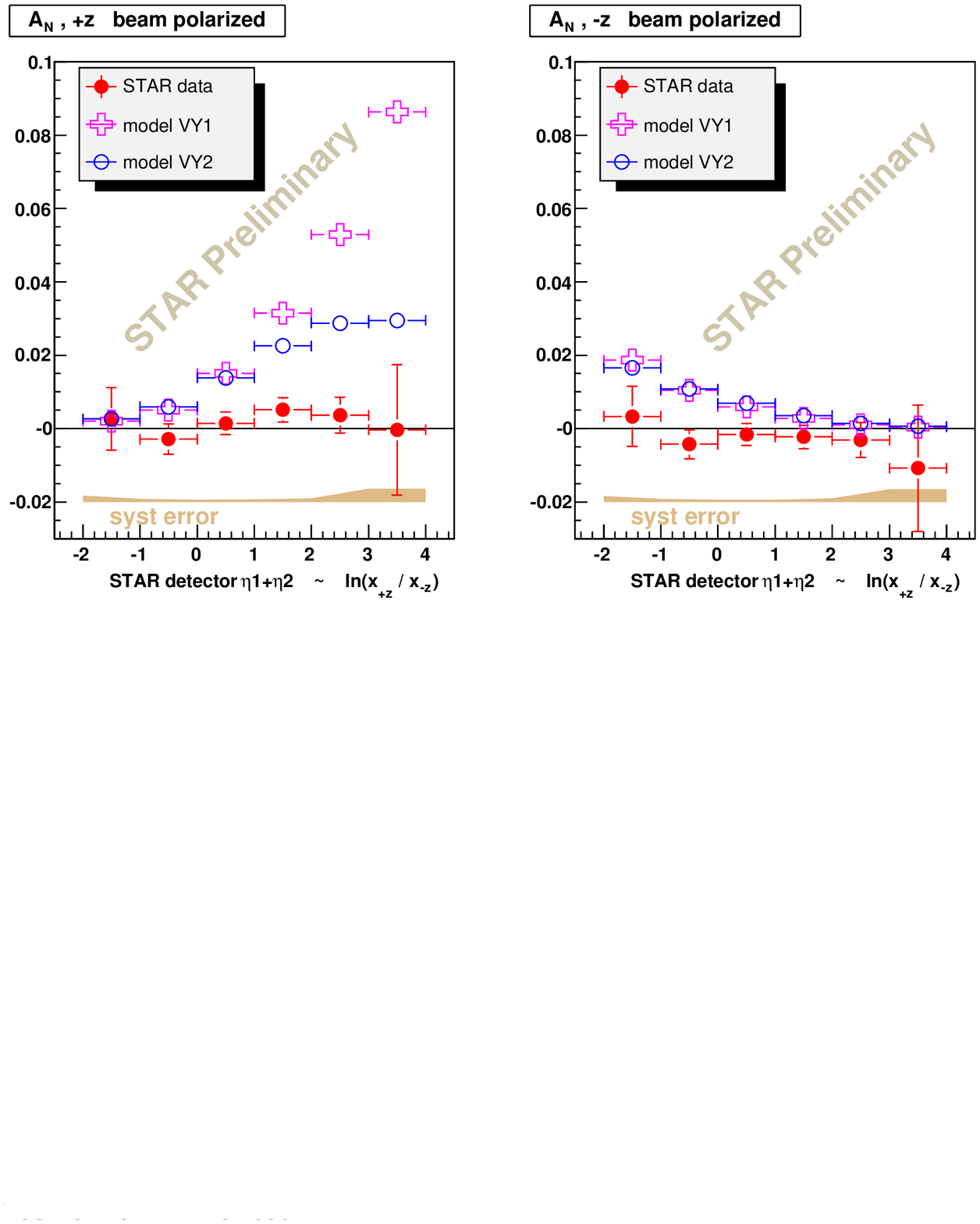}
 \caption{\label{fig:dijetSivers}\small Comparison of STAR results for
 $A^{+z}_N$ (left) and $A^{-z}_N$ (right) as function of $\eta_1 +\eta_2$ with
 predictions \cite{PRD72-054028}, based on two models of quark Sivers
 functions deduced from fits to HERMES SIDIS results
 \cite{hepex-0507013}.}
 \end{wrapfigure}
 access to the Sivers effect \cite{PRD72-054028}.
 Figure~\ref{fig:dijetSivers} shows preliminary results from 2006 data for
 $A_N$, for both beams, as a function of $\eta_1 +\eta_2$, where $\eta_1$
 and $\eta_2$ are the psuedorapidities for the two jets. The
 measurement was done over $\eta\in[-1,2]$ and for
 $|\zeta-180^\circ|\le68^\circ$. $A_N$ is extracted from the cross
 ratio of the spin-sorted $\zeta$ distribution in which ``left'' and
 ``right'' yields were deduced from $\zeta>180^\circ$ and
 $\zeta<180^\circ$, respectively. $A_N^{\pm z}$, shown in
 Figure~\ref{fig:dijetSivers}, are consistent with zero - much smaller
 than the Sivers asymetry that was observed in semi-inclusive deep
 inelastic scattering (SIDIS) by HERMES \cite{hepex-0507013}.\\

\medskip
\medskip

  The STAR collaboration uses the measurement of the longitudinal double
 spin asymmetry, $A_{LL}$, for inclusive jet and inclusive hadron
 production to constrain $\Delta g$. $A_{LL}$ is directly sensitive to the
 polarized gluon distribution function in the proton through gluon-gluon
 and gluon-quark sub-processes \cite{prd70-034010}.
  We reported the unpolarized inclusive jet cross section from data taken
 in 2003 and 2004 \cite{PRL97-252001}, and preliminary results for the
 unpolarized inclusive $\pi^0$ cross section \cite{hepex-0612004}, and
 both show satisfactory agreement with NLO pQCD calculations over many
 orders of magnitude. The agreement suggests the applicability of pQCD to
 describe the polarization observables in this kinematic region.

\medskip

 \begin{figure}[h]
 \begin{minipage}{10.8pc}
 \includegraphics[scale=0.4]{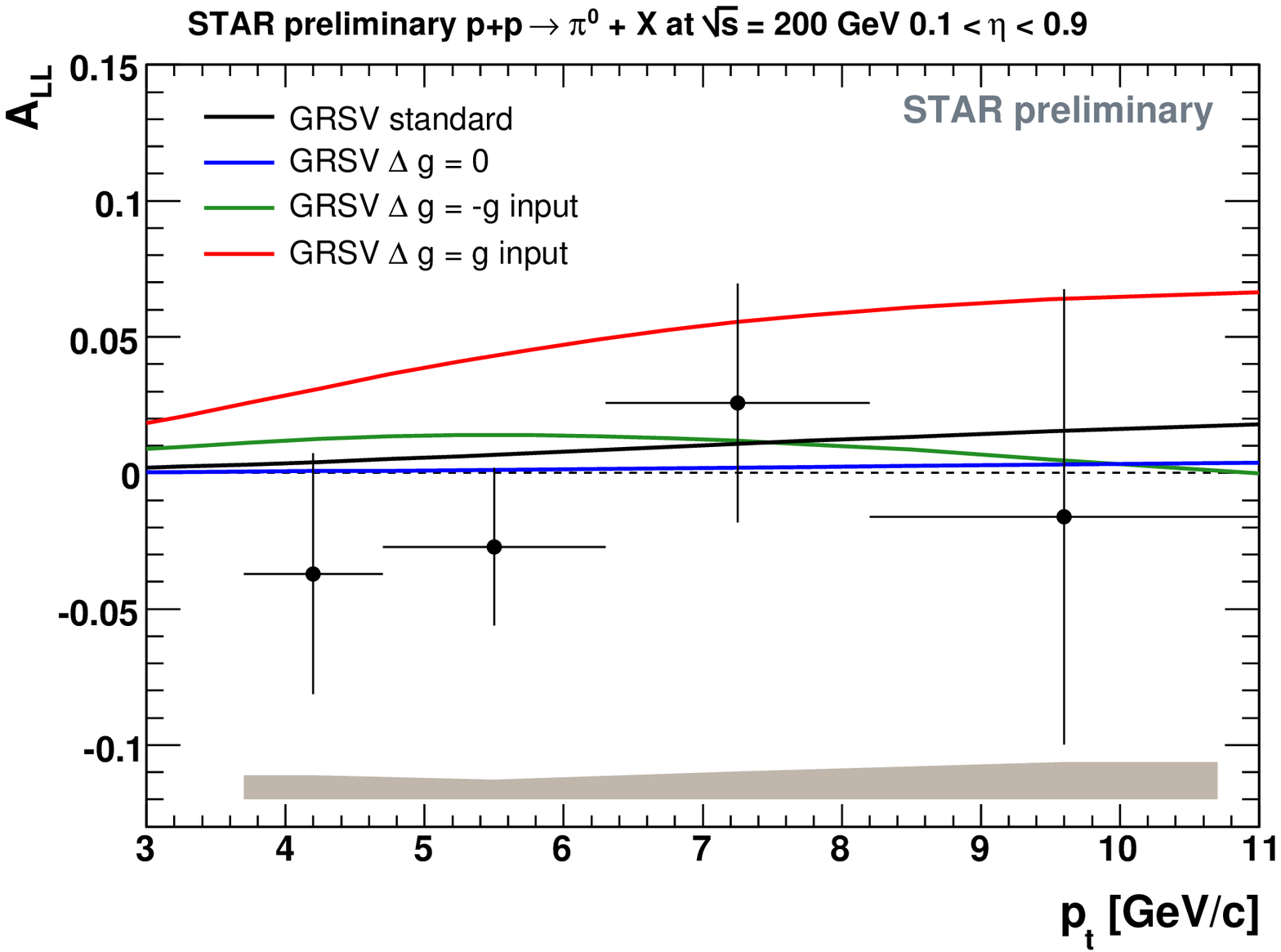}
 \end{minipage}\hspace{8pc} 
 \begin{minipage}{10pc}
 \includegraphics[scale=0.4]{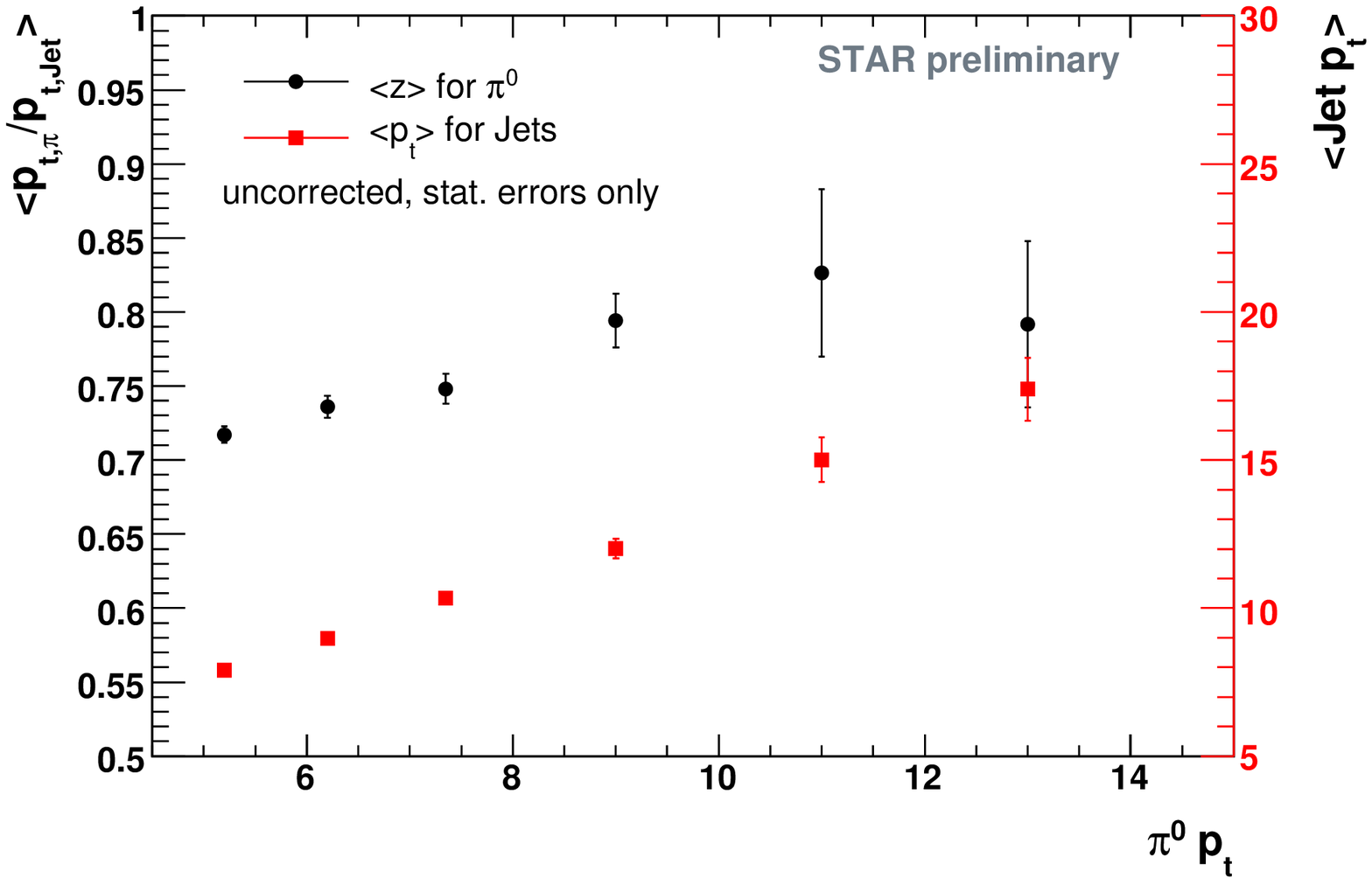}
 \end{minipage}
 \hspace{4pc}
 \caption{\label{fig:aLLpi0nz}\small Left panel : $A_{LL}$ for inclusive
 $\pi^\circ$ production with statistical errors and the systematic band.
 Right panel : Mean momentum fraction of the $\pi^\circ$ in their
 associated jet as a function of $p_T$ and the corresponding mean jet
 $p_T$. The data points are not corrected for jet acceptance or
 reconstruction efficiency.}
 \end{figure}
   In Figure~\ref{fig:aLLpi0nz}, the left panel, we present preliminary
 results of the first STAR $A_{LL}$  measurement of inclusive $\pi^\circ$
 production at mid-rapidity with statistical errors and the systematics
 band \cite{hepex-0612004}.
 The systematics band does not include the normalization uncertainty
 due to the polarization measurement. In addition to the GRSV standard
 curve, which is based on the best fit to DIS data, different $\Delta g$
 scenarios from maximally positive to maximally negative and passing zero
 $\Delta g$ are also shown in this and the following figures
 \cite{PRD67-054005,JHEP0207-012}.
 This measurement is consistent with previously reported $A_{LL}$ for
 inclusive jets \cite{PRL97-252001} in disfavoring the maximum positive
 $\Delta g$. The right panel shows the mean momentum fraction carried by
 the $\pi^\circ$ in its associated jet. This plot shows that the
 $\pi^\circ$ carries a very large fraction, 70\%, of its associated jet
 momentum. Corrections for jet acceptance and reconstruction efficiency
 may reduce this fraction by $\sim10\%$.

 \begin{figure}[h]
 \begin{minipage}{10pc}
 \includegraphics[scale=0.3]{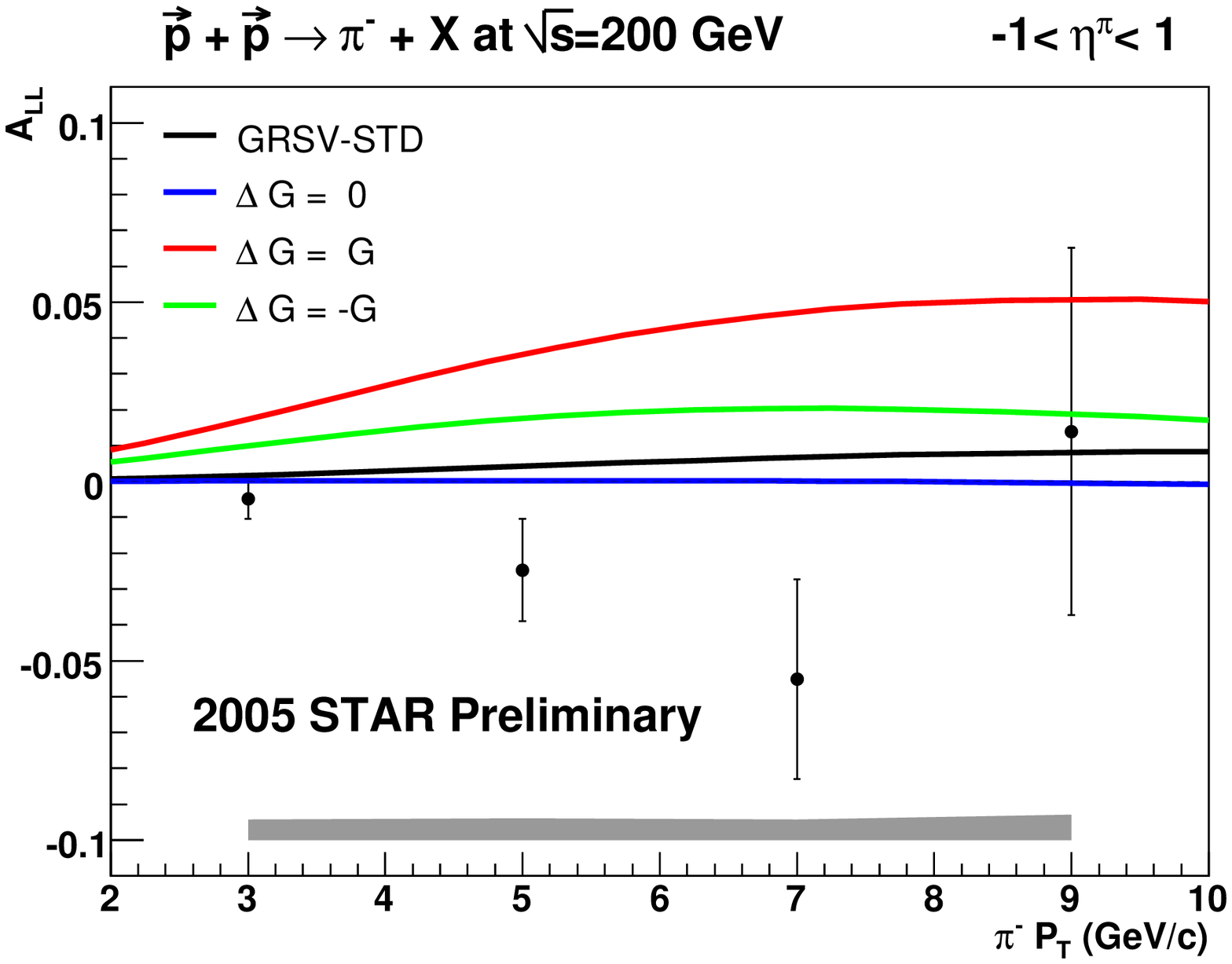}
 \end{minipage}\hspace{3pc} 
 \begin{minipage}{10pc}
 \includegraphics[scale=0.3]{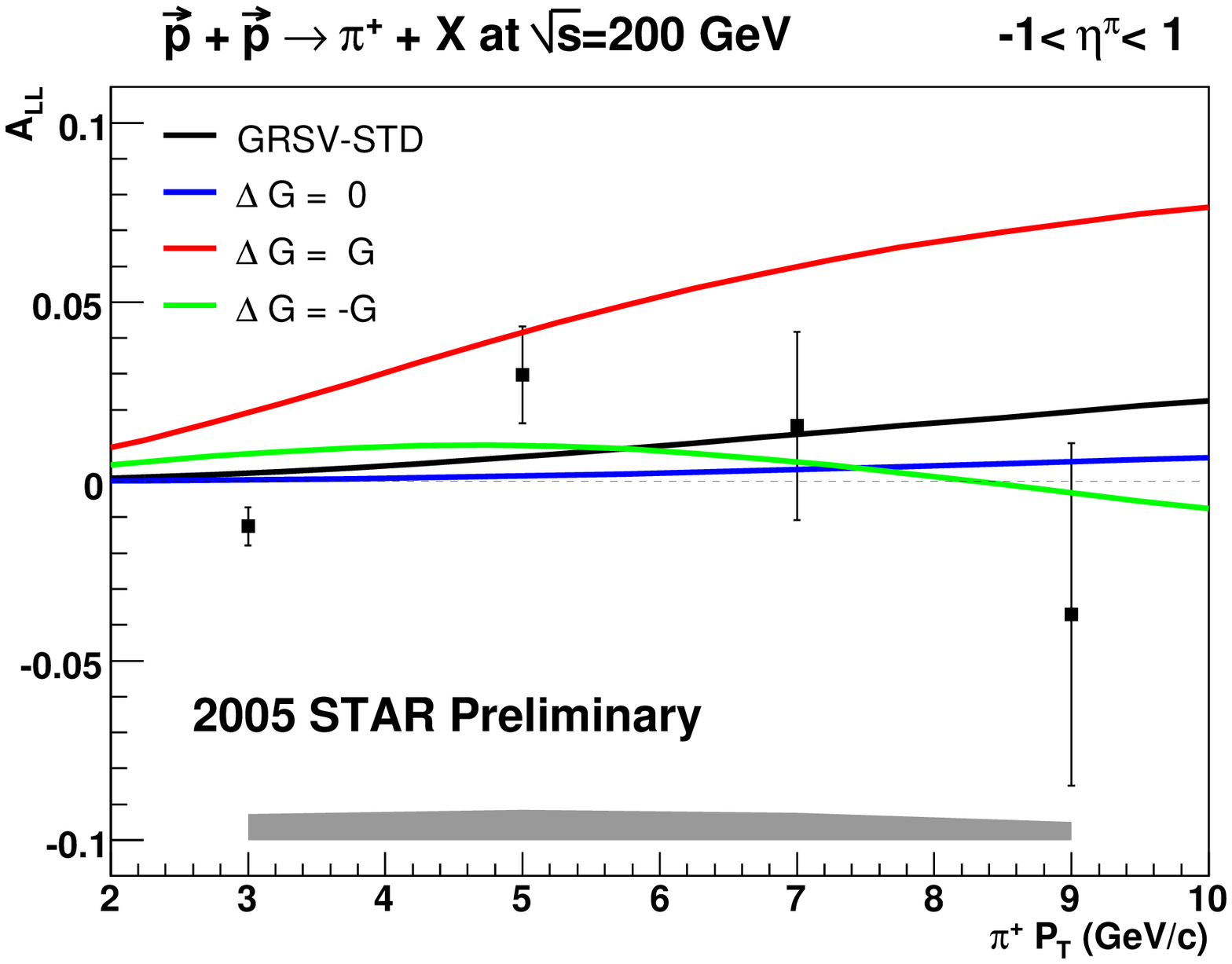}
 \end{minipage}
 \hspace{4pc}
 \begin{minipage}{14pc}
 \parbox[h]{4.4cm}{\caption{\label{fig:aLLplusmin}\small $A_{LL}$ for
 inclusive $\pi^-$ in the left panel and inclusive $\pi^+$ in the right
 one with statistical errors, and point-to-point systematic bands. The
 asymmetries are compared to theoretical predictions.}}
 \end{minipage}%
 \end{figure}

\medskip

 Preliminary results from 2005 of very promising $A_{LL}$ measurements of
 inclusive $\pi^+$ and $\pi^-$ are shown in Figure~\ref{fig:aLLplusmin}
 \cite{hepex-0612005}.
 The error bars are statistical errors, and the associated bands indicate
 the systematics, which don't include the scale uncertainty due to the
 polarization measurement. Several $\Delta g$ scenarios for these
 processes are also shown. The ordering of the
 measurements of $A_{LL}$ between $\pi^+$ and $\pi^-$ is sensitive to the
 sign of $\Delta g$. For now, this measurement is limited by statistics.
 We are currently working on 2006 data where we have higher luminosity and
 polarization.

 \begin{figure}[h]
   \begin{minipage}{17pc}
     \includegraphics[scale=0.5]{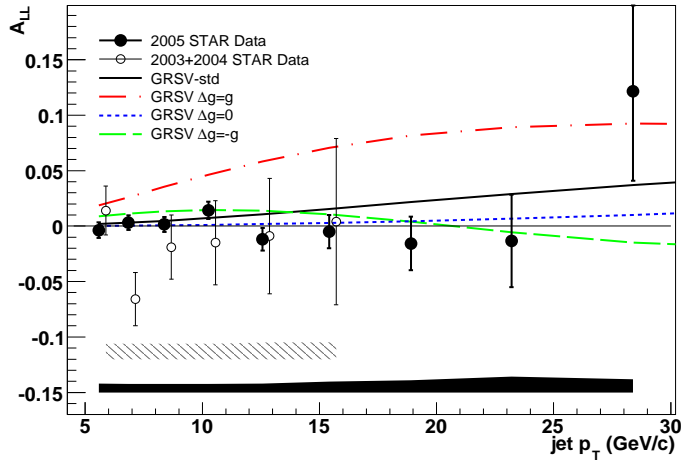}
   \end{minipage}\hspace{8.5pc}
   \begin{minipage}{22pc}
     \parbox[h]{5cm}{\caption{\label{fig:aLLjets}\small A$_{LL}$ for
 inclusive jet production at $\sqrt{s}=200$ GeV versus jet $p_T$ with
 statistical error bars. The bands indicate the systematic
 uncertainties, which do not include the scale uncertainty due to
 polarization measurement. The curves are theoretical calculations
 \cite{prd70-034010} based on NLO pQCD.}}
   \end{minipage}%
 \end{figure}

 Fig.~\ref{fig:aLLjets} shows $A_{LL}$ for inclusive jet production versus
 jet $p_T$. The solid points are preliminary results from 2005 data
 and the open circles are from 2003+2004 data analysis
 \cite{PRL97-252001}, both with statistical error bars. The dark and
 hatched bands show the systematic uncertainties for 2005 and combined
 2003 and 2004 data, respectively. These bands do not include scale
 uncertainties due to the beam polarization.
 Fig.~\ref{fig:aLLjets} also shows several GRSV NLO pQCD
 scenarios for inclusive jets. The data are consistent with zero or small
 gluon polarization and exclude the maximally positive gluon polarization
 scenario. In fact, these 2005 results go further in excluding much of the
 region in between maximally positive gluon polarization and DIS based fit
 scenarios, and thus put a tighter constraint on $\Delta g$.

\medskip

 In addition, the STAR collaboration studied longitudinal spin transfer of
 $\Lambda ( \bar\Lambda)$ in polarized p+p collisions\cite{xuProc} and
 inclusive $\pi^\circ$ production using the STAR Endcap Calorimeter
% \cite{jasonProc} as proof of principle measurements in 2005.
 as proof of principle measurements in 2005.

\medskip

   In summary, we reported on preliminary di-jet Sivers $A_N$ measurement
 from 2006, and all measured $A_N$ are statistically consistent with zero,
 several factors smaller than expected from SIDIS. In addition, we
 reported on 2006 $A_N$ measurements of forward $\pi^\circ$ with higher
 statistics and polarization, which show inconsistency with Twist-3 /
 Sivers calculations; $A_N$ does not decrease with $p_T$ in all $x_F$
 bins.

  In the longitudinal spin program, we reported on several preliminary
 $A_{LL}$ measurements of inclusive hadrons and jets at mid-rapidity from
 2005 data. The results are consistent in the region of kinematic overlap
 and provide improved constraints on the gluon spin contribution to the
 proton spin. They exclude large and positive gluon spin contributions.

\end{document}